\documentstyle[12pt]{article}
\def\href#1#2{{#2}}
\begin{document}
\begin{titlepage}
\begin{flushright}
hep-th/9908205 \\
TS-TH-99-1 \\
August 1999
\end{flushright}
\vspace{12pt}
\begin{center}
\LARGE
10 = 6 + 4 \\
\vspace{24pt}
\normalsize
Frank D. (Tony) Smith, Jr. \\
\vspace{6pt}
\footnotesize
e-mail: tsmith@innerx.net  \\
P. O. Box 370, Cartersville, GA 30120 USA \\
WWW URL: \\
http://www.innerx.net/personal/tsmith/TShome.html \\
\end{center}
\normalsize
\vspace{24pt}

\begin{abstract}

Some physics models have 10 dimensions that are usually decomposed into: \\
4 spacetime dimensions with local Lorentz $Spin(1,3)$ symmetry \\
plus \\
a 6-dimensional compact space related to internal symmetries. \\
\\
A possibly useful alternative decomposition is into: \\
6 spacetime dimensions with local Conformal symmetry \\
of the Conformal Group $C(1,3) = Spin(2,4) = SU(2,2)$ \\
plus \\
a 4-dimensional compact Internal Symmetry Space \\
that can be taken to be complex projective 2-space ${\bf{C}}P^{2}$ \\
which, since ${\bf{C}}P^{2} = SU(3) / U(2)$, \\
is a natural representation space for $SU(3)$ \\
and on which $U(2) = SU(2) \times U(1)$ can be represented naturally
by local action. \\

\end{abstract}
\normalsize
\end{titlepage}
\newpage
\setcounter{footnote}{0}
\setcounter{equation}{0}

\tableofcontents

\newpage

\section{Decomposition of 10 Dimensions}

Some physics models have 10 dimensions that are usually decomposed into:

\vspace{12pt}

4 spacetime dimensions with local Lorentz $Spin(1,3)$ symmetry

\vspace{12pt}

plus

\vspace{12pt}

a 6-dimensional compact space related to internal symmetries. \\

\vspace{12pt}

\vspace{12pt}

A possibly useful alternative decomposition is into:

\vspace{12pt}

6 spacetime dimensions with local
$C(1,3) = Spin(2,4) = SU(2,2)$ Conformal symmetry.

\vspace{12pt}

plus

\vspace{12pt}

a 4-dimensional compact Internal Symmetry Space.

\vspace{12pt}

\subsection{6-Dimensional Conformal spacetime}

\vspace{12pt}

Conformal symmetries and some of their physical applications
are described in the book of Barut and Raczka \cite{BR}.

\vspace{12pt}

The Conformal group $C(1,3)$ of Minkowski spacetime
is the group $SU(2,2) = Spin(2,4)$. As $Spin(2,4)$,
the Conformal group acts on a 6-dimensional (2,4)-space
that is related to the 6-dimensional ${\bf{C}}P^{3}$ space
of Penrose twistors \cite{PR}.

\vspace{12pt}

It is reasonable to consider the 6-dimensional Conformal space
as the spacetime in the dimensional decomposition
of 10-dimensional models
because Conformal symmetry is consistent with
such physics structures as: \\

Maxwell's equations of electromagnetism; \\

the quantum theoretical hydrogen atom; \\

the canonical Dirac Lagrangian for massive fermions,
as shown by Liu, Ma, and Hou \cite{LMH}; \\

gravity derived from the Conformal group using
the MacDowell-Mansouri mechanism,
as described by Mohapatra \cite{Moh}; \\

the Lie Sphere geometry of spacetime correlations; and \\

the Conformal physics model of I. E. Segal \cite{IES}. \\

\vspace{12pt}

\vspace{12pt}

\subsection{4-Dimensional Internal Symmetry Space}

\vspace{12pt}

An example of a possibly useful
4-dimensional compact Internal Symmetry Space
is complex projective 2-space ${\bf{C}}P^{2}$.

\vspace{12pt}

Since ${\bf{C}}P^{2} = SU(3) / U(2)$,
it is a natural representation space for $SU(3)$.

\vspace{12pt}

Further, $U(2) = SU(2) \times U(1)$ can be represented naturally
on ${\bf{C}}P^{2} = SU(3) / U(2)$ as a local action.

\vspace{12pt}

Therefore, all three of the gauge groups of the Standard Model
$SU(3) \times SU(2) \times U(1)$ can be represented
on the 4-dimensional compact Internal Symmetry Space
${\bf{C}}P^{2} = SU(3) / U(2)$.

\vspace{12pt}

\vspace{12pt}

The following section lists
some examples of physics models that have such 10-dimensional spaces:
Superstring theory; the Division Algebra model of Geoffrey Dixon;
and the $D_{4}-D_{5}-E_{6}-E_{7}$ physics model.

\vspace{12pt}

\newpage

\section{Superstrings, Dixon, and D4-D5-E6-E7}

\vspace{12pt}

\subsection{Superstrings}

\vspace{12pt}

The 10-dimensional space of Superstring theory is well known,
and described in many references, so I will not try to
summarize it here. One particularly current
and thorough reference is the 2-volume work of Polchinski
\cite{Pol}.

\vspace{12pt}

\subsection{Geoffrey Dixon's Division Algebra model}

\vspace{12pt}

Geoffrey Dixon, in his publications and website \cite{Dix},
considers the real division algebras: \\
the real numbers $\bf{R}$; \\
the complex numbers $\bf{C}$; \\
the quaternions $\bf{Q}$; and \\
the octonions $\bf{O}$. \\

\vspace{12pt}

Dixon then forms the tensor product
${\bf{T}} = {\bf{R}} \otimes {\bf{C}} \otimes {\bf{Q}} \otimes {\bf{O}}$
and considers the 64-real-dimensional space $\bf{T}$.

\vspace{12pt}

Then Dixon takes the left-adjoint actions
${\bf{T}_{L}} = {\bf{C}_{L}} \otimes {\bf{Q}_{L}} \otimes {\bf{O}_{L}}$,
and notes that ${\bf{T}_{L}}$ is isomorphic to
${\bf{C}}(16) = Cl(0,9) = {\bf{C}} \otimes Cl(0,8)$.

\vspace{12pt}

Then Dixon considers the algebra $\bf{T}$ to be
the spinor space of ${\bf{T}_{L}}$.

\vspace{12pt}

Then Dixon forms a matrix algebra ${\bf{T}}_{L}(2)$ as
the $2 \times 2$ matrices whose elements are in
the left-action adjoint matrix algebra ${\bf{T}_{L}}$
and notes that ${\bf{T}_{L}(2)}$ is isomorphic to
${\bf{C}}(32) = {\bf{C}} \otimes Cl(1,9)$.

\vspace{12pt}

Dixon describes the matrices ${\bf{T}_{L}(2)}$ as
having spinor space ${\bf{T}} \oplus {\bf{T}}$ and
${\bf{C}} \otimes Cl(1,9)$ as the Dirac algebra
of 10-dimensional (1,9)-space.

\vspace{12pt}

Dixon then describes leptons and quarks in
terms of reduction of the Dirac spinors
of the 10-dimensional (1,9)-space to
the Dirac spinors of a 4-dimensional (1,3)-spacetime.

\vspace{12pt}

The right-action adjoint matrix algebra ${\bf{T}_{R}}$ is not
the same as the left-action adjoint ${\bf{T}_{L}}$, because,
although ${\bf{C}_{R}} = {\bf{C}_{L}}$ and ${\bf{O}_{R}} = {\bf{O}_{L}}$,
it is a fact that  ${\bf{Q}_{R}} \neq {\bf{Q}_{L}}$
(they are isomorphic but not identical).

\vspace{12pt}

Since ${\bf{Q}_{R}} = {\bf{Q}}$, the part of the
matrix algebra ${\bf{T}_{R}}$ that differs from ${\bf{T}_{L}}$
is just ${\bf{Q}}$, and the different part of the $2 \times 2$
matrix algebra ${\bf{T}}_{R}(2)$ is just the $2 \times 2$
matrix algebra with quaternion entries ${\bf{Q}}(2)$.

\vspace{12pt}

In section 6.7 of his book \cite{Dix}, Dixon shows that
commutator closure of the set of traceless $2 \times 2$
matrices over the quaternions ${\bf{Q}}$,
which he denotes by $sl(2,{\bf{Q}})$, is
the Lie algebra of $Spin(1,5)$.

\vspace{12pt}

Since the Lie algebra $Spin(1,5)$ is just
the Lie algebra of the Conformal group $C(1,3) = Spin(2,4) = SU(2,2)$
with a different signature,

\vspace{12pt}

I conjecture that it might be useful to consider
the spacetime part of Dixon's 10-dimensional (1,9)-space
to be
the 6-dimensional (1,5)-spacetime of $Spin(1,5)$.

\vspace{12pt}

That would leave a 4-dimensional (0,4)-space to
be used as an Internal Symmetry Space.

\vspace{12pt}

\vspace{12pt}

\subsection{the D4-D5-E6-E7 model}

\vspace{12pt}

The $D_{5}$ Lie algebra of the $D_{4}-D_{5}-E_{6}-E_{7}$ physics model
corresponds (with Conformal signature) to the
Lie algebra $Spin(2,8)$ of the Clifford algebra $Cl(2,8)$
whose vector space is 10-dimensional. \\

As the $D_{4}-D_{5}-E_{6}-E_{7}$ physics model
is described on the web \cite{TS}, I will not try
to summarize it here.

\newpage

\section{Acknowledgements}

\vspace{12pt}

The idea of 6-dimenisonal spacetime with Conformal symmetry
was motivated by the works of I. E. Segal \cite{IES} and
by e-mail conversations with Robert Neil Boyd.

\vspace{12pt}

The idea of 4-dimensional Internal Symmetry Space
was motivated by Cayley calibrations of octonions
\cite{Har} and
by e-mail conversations with Matti Pitkanen.

\end{document}